\documentclass[10pt,twocolumn]{article}

\usepackage[a4paper,margin=0.75in]{geometry}
\usepackage[T1]{fontenc}
\usepackage[utf8]{inputenc}
\usepackage{lmodern}
\usepackage{microtype}
\usepackage{booktabs}
\usepackage{longtable}
\usepackage{array}
\usepackage{xcolor}
\usepackage{graphicx}
\usepackage{listings}
\usepackage{hyperref}
\usepackage{enumitem}
\usepackage{caption}
\usepackage{needspace}
\usepackage{amsmath}
\usepackage{amssymb}
\usepackage{pgfplots}
\usepackage{placeins}
\pgfplotsset{compat=1.18}
\usepackage{tikz}
\usetikzlibrary{arrows.meta,positioning,fit,backgrounds}
\usepackage{dblfloatfix}
\hypersetup{
  colorlinks=true,
  linkcolor=blue!50!black,
  citecolor=blue!50!black,
  urlcolor=blue!50!black
}

\lstdefinelanguage{CPL}{
  morekeywords={start,exit,function,container,return,if,else,while,loop,switch,case,default,glob,ro,dref,ref,ptr,lis,break,extern,from,import,syscall,asm,as,f64,f32,i64,i32,i16,i8,u64,u32,u16,u8,i0,str,arr,not,neg,poparg,sizeof,section,align,line,include,define,undef,ifdef,ifndef,endif,self,like_c},
  sensitive=true,
  morecomment=[s]{:/}{/:},
  morestring=[b]",
}

\lstdefinelanguage{PTRN}{
  morekeywords={mov,add,sub,mul,div,mod,shl,shr,and,or,xor,test,cmp,jmp,mklb,delete,DELETE_LINE,const,obj,reg},
  sensitive=true,
  morecomment=[l]{;},
  morestring=[b]",
}

\title{CPL: A Compact C-like Systems Language with Explicit Low-Level Control}
\author{
  \normalsize Nikolai Fot \and \normalsize Alexander Vinarsky \\
  \normalsize Compiler Technology Department, \\
  \normalsize Ivannikov Institute for System Programming of the Russian Academy of Sciences (ISP RAS)
}
\date{}

\begin{document}
\maketitle

\begin{abstract}
\par This paper presents Cordell Programming Language (CPL), a compact C-like systems language that retains C's direct access to memory, layout, and machine interfaces while experimenting with a smaller grammar and selected conveniences from newer languages. Also this paper studies whether C-like are more convenient to use for compiler experiments than modern approaches and paradigms. While the language and compiler provide primitive values, pointers, arrays, containers with methods, unions, generic functions, overloads, entry‑point and section control, system calls, and inline assembly, they do not provide high‑level constructs such as classes, built‑in methods, a standard library, or memory protection. The article describes the language design, compiler pipeline, target backends, static‑analysis architecture, and OS‑facing use cases, then evaluates the prototype backend with reproducible x86\_64 and i386 microbenchmarks against C compiler baselines. The obtained results suggest that the compiler can produce code comparable to that produced by production compilers such as GCC and Clang, as well as by small compilers such as TinyC and SmallerC.
\end{abstract}

\section{Introduction}
\par Systems programming languages expose machine-level details while attempting to keep programs tractable. Mature languages such as C~\cite{ritchie1978c,kernighan1988c}, C++~\cite{stroustrup2007cpp,stroustrup2013cpp}, Rust~\cite{klabnik2023rust,jung2018rustbelt}, and Zig~\cite{ziglangref} provide large ecosystems, but the full syntactic and semantic complexity makes them expensive targets for compiler experiments. Conversely, many teaching compilers are intentionally small but do not expose enough low-level mechanisms for studying system-oriented code generation, platform-specific interfaces, and optimization under realistic constraints.
\par To address this issue building a new small compiler (and a language) is a promising approach to create a convenient testbed for learning and experiments. A new compiler provides more opportunities for experimentation, given the nature of optimizations, and a new language can reflect modern syntax common among languages such as Rust, C, C++, and Zig. The motivation for a new grammar lays in a fact, that a faithful C-like, Rust-like or Zig-like subset would either require a substantially more complex parser and semantic model, or create misleading expectations that the rest of the language should also be supported. The design instead favors a compact parser-friendly grammar, explicit syntax for functions and pointer-like operations, and compiler-supported low-level constructs.
\par The central research problem is the trade-off between implementation compactness and systems-level adequacy. Reducing a language and compiler makes experimentation and inspection easier, but can remove the layout controls, ABI mechanisms, analyses, and backend structure needed for realistic low-level programs. This article studies CPL as a C-like language design that keeps the directness of C while experimenting with a smaller grammar and selected modern conveniences.

\section{Language Overview}
\par The language remains close to assembly and C by exposing pointers, explicit dereferencing, C-like structures, direct system calls, inline assembly, manual entry points, and annotations for placement and control. It deliberately avoids reproducing the full C grammar and semantics. The compiler includes enough phase separation to study SSA-form construction~\cite{cytron1991ssa}, diagnostics, scalar optimization, low-level selection, register allocation, and peephole rewriting.

\subsection{Program Entry}
\par A program can use a \texttt{start} block as its static entry point. The entry block does not return a value with \texttt{return}; it must terminate through \texttt{exit}. An ordinary function can also become an entry point through the \texttt{@[entry]} annotation. The \texttt{start} structure was chosen as the basic name of a function, which is an entry point during the standard linker process with the \texttt{ld} tool on Linux. In most cases, it is more convenient to use the \texttt{@[entry]} annotation rather than explicit \texttt{start} function.

\FloatBarrier
\begin{lstlisting}[language=CPL,caption={Minimal CPL entry point.}]
:/ The final name depends on the architecture /:
start() {
    exit 0;
}

:/ 'main' is the name in the final assembly file /:
@[entry("main")] function main() {
}
\end{lstlisting}
\FloatBarrier

\par The entry point may accept arguments, such as \texttt{argc} and \texttt{argv}, and can be annotated for platform-specific entry behavior. The \texttt{naked} annotation disables default entry and exit routines, which is useful for system-level code that must manage its own prologue and epilogue. On i386 this also affects how stack arguments are addressed, as discussed in Section~\ref{sec:i386-abi}.

\subsection{Types}
\par CPL uses permissive static typing. Variables do not dynamically change type, widening conversions may be inserted implicitly, and narrowing conversions require an explicit \texttt{as} cast. The primitive integer and floating-point types include \texttt{i8}, \texttt{u8}, \texttt{i16}, \texttt{u16}, \texttt{i32}, \texttt{u32}, \texttt{i64}, \texttt{u64}, \texttt{f32}, \texttt{f64}, and the void-like function return type \texttt{i0}. Boolean-like logic follows the C convention: zero is false and non-zero is true. The sizes can be seen in the Table~\ref{tab:type_sizes}. The largest type size depends on the target system.

\begin{table}[t]
    \centering
    \footnotesize
    \caption{Type size on x86\_64 GNU.}
    \label{tab:type_sizes}
    \begin{tabular}{p{0.24\linewidth}p{0.20\linewidth}p{0.32\linewidth}}
    \toprule
    Type          & Size (bytes) \\
    \midrule
    u8, i8        & 1             \\
    u16, i16      & 2             \\
    u32, i32, f32 & 4             \\
    u64, i64, f64 & 8             \\
    \bottomrule
    \end{tabular}
\end{table}

\subsection{Pointers, Strings, and Arrays}
\par The language exposes pointer types through the \texttt{ptr} modifier. In contrast to the asterisk used in C, the explicit \texttt{ptr} keyword makes the pointers more distinct and imposes the uniform declaration style. The \texttt{ref} keyword obtains a pointer to an object or string literal, and \texttt{dref} reads or writes through a pointer.

\begin{lstlisting}[language=CPL,caption={Pointer operations in CPL.}]
i32 x = 123;
ptr i32 p = ref x;
:/ While C supports int* p; int *p; int * p; /:
i32 y = dref p;
dref p = y + 1;
\end{lstlisting}

\par Strings and arrays are distinct built-in containers. \texttt{arr[0, i8]} allocates a terminated byte sequence, while \texttt{arr} declares a fixed-size array. String literals placed independently in code reside in a read-only section-like area and must be referenced explicitly.

\begin{lstlisting}[language=CPL,caption={Strings and arrays.}]
arr msg[0, i8] = "Hello world!"; :/ Compiler sets the size by itself, which means we can provide '0' /:
arr xs[4, i32] = { 1, 2, 3, 4 }; :/ We can set the exact size for an array /:
ptr i8 p = ref "Hello, World!\n";
\end{lstlisting}

\subsection{Containers}

\par CPL containers are lightweight C-like aggregate types. They group fields of different types in one named layout and are intended for explicit systems code rather than for object-oriented programming.

\begin{lstlisting}[language=CPL,caption={A simple CPL container.}]
container pair {
    i32 first;
    i32 second;
}
\end{lstlisting}

\par By default, container fields use the target architecture's default maximum alignment policy. A container can override this with the \texttt{align} annotation. For example, \texttt{@[align(1)]} requests a packed layout, which is useful when a byte-exact representation is required.

\begin{lstlisting}[language=CPL,caption={Packed container layout.}]
@[align(1)]
container packed_value {
    i8  a;
    i16 b;
    i32 c;
} :/ packed, occupies 7 bytes on the stack /:
\end{lstlisting}

\par In addition to fields, a container may declare functions. These functions are ordinary CPL functions attached syntactically to the container definition. They do not make the container a class, and they do not introduce implicit constructors or destructors.

\begin{lstlisting}[language=CPL,caption={Container function.}]
container counter {
    @[inline(always)]
    function default_value() -> i32;
}

function counter::default_value() -> i32 {
    return 100;
}

counter c;
c.default_value();
\end{lstlisting}

\par The \texttt{self} annotation marks a container function whose first explicit argument is the receiver pointer. Calls to such functions may use method-like syntax, and the parser can pass the container pointer automatically. This is purely syntactic sugar only: a \texttt{self} function must still declare its receiver argument explicitly.

\begin{lstlisting}[language=CPL,caption={A self container function.}]
container counter {
    i32 val;

    @[self]
    function init(ptr counter self) -> i0 {
        self.val = 100;
    }
}

counter c;
c.init();
c.val; :/ 100 /:
\end{lstlisting}

\par Containers should therefore be treated as modified C structures with optional attached functions. Initialization, cleanup, and ownership-like behavior must be written explicitly by the programmer.
\par Containers may also represent unions. With \texttt{@[union]}, all fields share offset zero and the aggregate size is determined by the largest field together with the selected alignment policy. The annotation composes with \texttt{@[align]} and \texttt{@[like\_c]}.

\begin{lstlisting}[language=CPL,caption={Union container with C-compatible layout.}]
@[like_c]
@[union]
container device_value {
    i8  size_8;
    i16 size_16;
    i32 size_32;
    i64 size_64;
}
\end{lstlisting}

\par Container values can be array elements, and array element types may themselves be arrays. These forms permit arrays of records and statically sized matrices without introducing a separate aggregate system.

\begin{lstlisting}[language=CPL,caption={Arrays of containers and nested arrays.}]
container string_view {
    ptr i8 data;
    i32    length;
}

arr views[10, string_view];
arr matrix[10, arr[10, i32]];
\end{lstlisting}

\par A container method may be declared in the container and defined separately with the \texttt{::} qualifier. This form avoids emitting the same implementation in every translation unit while retaining container-qualified lookup.

\begin{lstlisting}[language=CPL,caption={Out-of-container method definition.}]
container math {
    glob function add(i32 a, i32 b) -> i32;
}

function math::add(i32 a, i32 b) -> i32 {
    return a + b;
}
\end{lstlisting}

\subsection{Control Flow}
\par CPL provides \texttt{if}, \texttt{while}, \texttt{loop}, and \texttt{switch}. A semicolon separates the condition from the body in conditional constructs. The \texttt{loop} construct represents an infinite loop unless terminated with \texttt{break} or annotated with \texttt{@[counter(n)]}. The \texttt{switch} construct supports fallthrough by default, while \texttt{@[no\_fall]} inserts break-like behavior into cases.

\begin{lstlisting}[language=CPL, caption={Control flow.}]
if x > 0; x -= 1;
else      x = 0;

@[counter(10)] loop {
    x += 1;
}
\end{lstlisting}

\subsection{Functions}
\par Functions are declared with the \texttt{function} keyword. CPL supports prototypes, default arguments, local functions, function overloading with restrictions, generic functions, function pointers, lambdas without capturing closures, and variadic arguments. Global functions use \texttt{glob} when their symbol name must be preserved for external linkage.

\begin{lstlisting}[language=CPL, caption={Function forms.}]
function add(i32 a, i32 b = 1) -> i32 {
    return a + b;
}

glob function exported(i32 x) -> i32;
\end{lstlisting}

\subsection{System-Level Facilities}
\par CPL provides two built-in low-level mechanisms: \texttt{syscall} and \texttt{asm}. A \texttt{syscall} expression is target-dependent and accepts a platform-specific argument list. Inline assembly supports argument substitution through numbered placeholders.

\begin{lstlisting}[language=CPL, caption={Inline assembly with placeholders.}]
i32 a = 0;
i32 ret;
asm(a, ret) {
    "push rax",
    "mov rax, %0",
    "syscall",
    "mov %1, rax",
    "pop rax"
}
\end{lstlisting}

\par Inline assembly is intentionally powerful and fragile. It is copied with minimal changes into the generated assembly after placeholder substitution, therefore, the programmer must preserve registers and match the selected target syntax. The compiler does not optimize inside inline assembly blocks.

\subsection{Annotations}

\par Annotations extend the small grammar with low-level behavior. Table~\ref{tab:annotations} summarizes the system-oriented annotations available in the implementation. The maturity column separates stable annotations from annotations whose behavior is target-dependent or experimental.

\begin{table*}[t]
\centering
\footnotesize
\caption{Selected CPL annotations.}
\label{tab:annotations}
\begin{tabular}{p{0.12\linewidth}p{0.38\linewidth}p{0.22\linewidth}p{0.12\linewidth}}
\toprule
Annotation & Purpose & Example & Maturity \\
\midrule
\texttt{naked}             & Disable default entry and exit routines. & \texttt{@[naked] start()} & stable \\
\texttt{align}             & Align a local or global object. & \texttt{@[align(16)] glob i32 a;} & stable \\
\texttt{section}           & Place code or data in a target section. & \texttt{@[section(".data")]} & stable \\
\texttt{nosection}         & Emit a declaration without an explicit section directive. & \texttt{@[nosection] glob function f()} & stable \\
\texttt{entry}             & Mark a function as an entry point. & \texttt{@[entry("\_start")]} & stable \\
\texttt{no\_fall}          & Insert break-like behavior in switch cases. & \texttt{@[no\_fall] switch x;} & stable \\
\texttt{straight}          & Use linear switch selection. & \texttt{@[straight] switch x;} & stable \\
\texttt{counter}           & Generate a counted loop. & \texttt{@[counter(100)] loop} & stable \\
\texttt{hot}/\texttt{cold} & Influence branch layout. & \texttt{@[cold] if cond;} & stable \\
\texttt{register}          & Bind a primitive variable to a register index. & \texttt{@[register(RAX)] i32 x;} & stable \\
\texttt{self}              & Mark a container method whose explicit receiver pointer can be passed through method-like syntax. & \texttt{@[self] function init(ptr T self)} & stable \\
\texttt{poparg}            & Influence argument popping at a low-level call boundary. & \texttt{@[poparg]} & stable \\
\texttt{like\_c}           & Request C-ABI-like padding and layout behavior for a container. & \texttt{@[like\_c] container T \{...\}} & stable \\
\texttt{union}             & Give all fields of a container offset zero. & \texttt{@[union] container U \{...\}} & implemented \\
\texttt{abi}               & Mark an external function declaration as ABI-compatible. & \texttt{@[abi] extern function f(...)} & implemented \\
\texttt{weak}              & Emit a weak linker symbol for a function. & \texttt{@[weak] function f()} & implemented \\
\texttt{only\_body}        & Emit only the lowered body of a function, without its symbol label or ordinary wrapper. & \texttt{@[only\_body] function h()} & experimental \\
\bottomrule
\end{tabular}
\end{table*}

\par These annotations expose system-level control without introducing an object-oriented runtime model. The \texttt{like\_c} annotation is relevant for containers that cross a C ABI boundary because it specifies C-compatible padding assumptions. The \texttt{section} annotation accepts an optional alignment argument, for example \texttt{@[section(".bss", 16)]}; this is distinct from aligning an individual declaration because the emitted section directive itself carries the alignment constraint. The \texttt{only\_body} annotation supports target directives or instruction sequences that must appear without a generated function label, while preserving the function body as an optimizable compiler object before emission.

\subsection{Core Static and Dynamic Model}
\label{sec:formal-core}

\par The complete language is defined by the implementation, but the safety-relevant core used in this article can be stated independently. Let primitive types be \(b \in \{\texttt{i8},\ldots,\texttt{u64},\texttt{f32},\texttt{f64},\texttt{i0}\}\), and let
\[
\tau ::= b \mid \texttt{ptr}\ \tau \mid \texttt{arr}[n,\tau] \mid C,
\]
where \(C\) is a container type. A typing environment \(\Gamma\) maps identifiers to types, a container environment \(\Delta\) maps field pairs \(C.f\) to field types, and \(\Gamma \vdash e : \tau\) denotes expression typing.

\begin{align*}
\frac{\Gamma \vdash x : \tau}
     {\Gamma \vdash \texttt{ref}\ x : \texttt{ptr}\ \tau}
&\qquad
\frac{\Gamma \vdash e : \texttt{ptr}\ \tau}
     {\Gamma \vdash \texttt{dref}\ e : \tau},
\\[0.5em]
\frac{\Gamma \vdash e : C \qquad \Delta(C,f)=\tau}
     {\Gamma \vdash e.f : \tau}.&
\end{align*}

\par Widening numeric conversions may be inserted by the compiler; narrowing conversions require an explicit \texttt{as} expression. Pointer dereference is type-correct when its operand has a pointer type, but typing does not establish allocation validity or non-nullness. This distinction motivates the separate SSA/SMT diagnostic layer.

\par A small-step memory model uses a store \(\sigma\) from addresses to typed values and locations \(\ell\):
\[
\langle \sigma,\texttt{ref}\ x\rangle \rightarrow
\langle \sigma,\ell_x\rangle,
\qquad
\frac{\sigma(\ell)=v}
     {\langle \sigma,\texttt{dref}\ \ell\rangle \rightarrow
      \langle \sigma,v\rangle}.
\]
For assignment through a pointer,
\[
\langle \sigma,\texttt{dref}\ \ell = v\rangle \rightarrow
\langle \sigma[\ell \mapsto v],v\rangle.
\]
Container field access is computed by adding the target-dependent field offset to the base location; union fields have a zero offset. Function calls evaluate arguments, establish parameter bindings, execute the body, and yield the explicit return value. Inline assembly, syscalls, and target annotations are modeled as externally defined transitions because their behavior depends on the selected ABI and machine. These rules define the article's core reasoning model, not a mechanized semantics or a proof that every compiler pass preserves it.

\section{Compiler Architecture}

\subsection{Pipeline}

\par
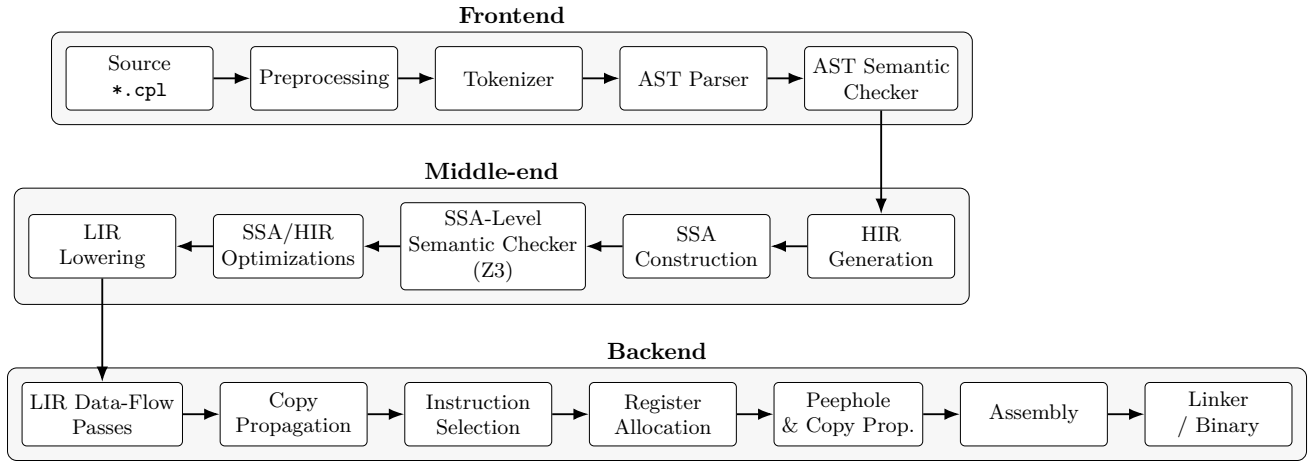
\begin{figure*}[t]
\centering
\resizebox{\textwidth}{!}{%
\begin{tikzpicture}[
    >=Latex,
    stage/.style={draw, rounded corners=2pt, align=center, minimum width=2.2cm, minimum height=0.95cm, fill=white, font=\small},
    group/.style={draw, rounded corners=4pt, inner sep=6pt, fill=black!3},
    arrow/.style={-Latex, thick}
]

% Row 1
\node[stage] (src) {Source\\\texttt{*.cpl}};
\node[stage, right=0.55cm of src] (prep) {Preprocessing};
\node[stage, right=0.55cm of prep] (tok) {Tokenizer};
\node[stage, right=0.55cm of tok] (ast) {AST Parser};
\node[stage, right=0.55cm of ast] (astsem) {AST Semantic\\Checker};

% Row 2
\node[stage, below=1.55cm of astsem] (hir) {HIR\\Generation};
\node[stage, left=0.55cm of hir] (ssa) {SSA\\Construction};
\node[stage, left=0.55cm of ssa] (ssasem) {SSA-Level\\Semantic Checker\\(Z3)};
\node[stage, left=0.55cm of ssasem] (opt) {SSA/HIR\\Optimizations};
\node[stage, left=0.55cm of opt] (hlir) {LIR\\Lowering};

% Row 3
\node[stage, below=1.55cm of hlir] (sched) {LIR Data-Flow\\Passes};
\node[stage, right=0.55cm of sched] (copy1) {Copy\\Propagation};
\node[stage, right=0.55cm of copy1] (isel) {Instruction\\Selection};
\node[stage, right=0.55cm of isel] (ra) {Register\\Allocation};
\node[stage, right=0.55cm of ra] (peephole) {Peephole \\ \& Copy Prop.};
\node[stage, right=0.55cm of peephole] (asm) {Assembly};
\node[stage, right=0.55cm of asm] (link) {Linker\\/ Binary};

% Group boxes
\begin{scope}[on background layer]
  \node[group, fit=(src)(prep)(tok)(ast)(astsem), label={[font=\bfseries]above:Frontend}] {};
  \node[group, fit=(hir)(ssa)(ssasem)(opt)(hlir), label={[font=\bfseries]above:Middle-end}] {};
  \node[group, fit=(sched)(copy1)(isel)(ra)(peephole)(asm)(link), label={[font=\bfseries]above:Backend}] {};
\end{scope}

% Arrows row 1
\draw[arrow] (src) -- (prep);
\draw[arrow] (prep) -- (tok);
\draw[arrow] (tok) -- (ast);
\draw[arrow] (ast) -- (astsem);
% inter-row
\draw[arrow] (astsem.south) |- ++(0,-0.42) -| (hir.north);
% row 2 reverse direction visually right-to-left
\draw[arrow] (hir) -- (ssa);
\draw[arrow] (ssa) -- (ssasem);
\draw[arrow] (ssasem) -- (opt);
\draw[arrow] (opt) -- (hlir);
% inter-row
\draw[arrow] (hlir.south) |- ++(0,-0.42) -| (sched.north);
% row 3
\draw[arrow] (sched) -- (copy1);
\draw[arrow] (copy1) -- (isel);
\draw[arrow] (isel) -- (ra);
\draw[arrow] (ra) -- (peephole);
\draw[arrow] (peephole) -- (asm);
\draw[arrow] (asm) -- (link);
\end{tikzpicture}%
}
\caption{Graphical overview of the CPL compiler architecture. The pipeline is arranged in three phases to preserve readability in a two-column layout while retaining the execution order.}
\label{fig:compiler-pipeline}
\end{figure*}

 Figure~\ref{fig:compiler-pipeline} presents the compiler architecture. The frontend performs preprocessing, tokenization, AST construction, and early semantic checking. The middle-end constructs HIR, converts it to SSA~\cite{cytron1991ssa}, runs SSA-level semantic checking with a Z3-backed symbolic layer~\cite{deMoura2008z3}, and applies HIR-level optimizations. The backend lowers the program to LIR, performs LIR data-flow and copy-propagation passes, selects target-sensitive instructions, allocates registers, applies late peephole cleanup, and finally emits assembly for an external assembler/linker toolchain.

\par This organization deliberately separates source-level analysis, SSA construction, symbolic checking, high-level optimization, lowering, instruction selection, register allocation, and late cleanup. The separation makes the implementation suitable for code generation, as well as for experiments with IR design, pass placement, and target-specific transformations.

\subsection{Worked Example}

\par Listing~\ref{lst:example-cpl} presents a minimal program used to illustrate the compiler forms.

\begin{lstlisting}[language=CPL,caption={Input CPL program.},label={lst:example-cpl}]
start() {
    i32 a = 1;
    i32 b = 2;
    exit a + b;
}
\end{lstlisting}

\par The corresponding HIR uses stack variables with an \texttt{s} suffix and temporaries with a \texttt{t} suffix.

\begin{lstlisting}[caption={High IR (HIR) form for Listing~\ref{lst:example-cpl}.}]
fn _main0() {
    i32s %0 = alloc;
    i32t %2 = i8n 1 as i32;
    i32s %0 = i32t %2;
    i32s %1 = alloc;
    i32t %3 = i8n 2 as i32;
    i32s %1 = i32t %3;
    i32t %5 = i32s %0 + i32s %1;
    u8t %4  = i32t %5 as u8;
    exit u8t %4;
}
\end{lstlisting}

\par After lowering to LIR, the program is represented as a basic-block sequence.

\begin{lstlisting}[caption={LIR form before target-sensitive selection.}]
BB1: start {
    %2 = $1 as i32;
    %0 = %2;
    %3 = $2 as i32;
    %1 = %3;
    %5 = %0 + %1;
    %4 = %5 as u8;
    exit %4;
    }
BB2: send
\end{lstlisting}

\par A later selected LIR form contains target-dependent register choices.

\begin{lstlisting}[caption={Selected LIR form.}]
BB1: start
rcx = $1;
rcx = rcx;
rdx = $2;
rdx = rdx;
rax = rcx;
rax = rax + rdx;
rcx = rax;
rcx = rcx;
rdi = rcx;
exit rdi;
BB2: send
\end{lstlisting}

\par The final optimized Mach-O assembly for this example collapses the computation to a short sequence.

\begin{lstlisting}[caption={Generated assembly.}]
section .text
global _main
_main:
mov al, 1
add al, 2
mov dil, al
mov rax, 0x2000001 ; Exit syscall on MACHO64
syscall
\end{lstlisting}

\subsection{Backends and Target Status}

\par The implemented backends are x86\_64 Mach-O/NASM, x86\_64 GNU/Linux NASM, and i386 GNU/Linux NASM. Each path includes instruction selection, memory selection, caller-saving logic, and assembly generation. Mach-O is the default configuration and has the broadest test coverage. Linux x86\_64 and i386 are exercised by dedicated compiler and assembly-generation tests; the i386 path additionally supports kernel-oriented examples. Other architectures and system options exposed by configuration code are outside the evaluated target set. The compiler can invoke \texttt{ld}, \texttt{clang}, or \texttt{gcc} for linking.

\begin{table}[t]
\centering
\footnotesize
\caption{Backend maturity in the CPL compiler prototype.}
\label{tab:backend-maturity}
\begin{tabular}{p{0.24\linewidth}p{0.20\linewidth}p{0.32\linewidth}}
\toprule
Target        & Assembler format & Status                       \\
\midrule
x86-64 Mach-O & \texttt{macho64} & default; broadest coverage   \\
x86-64 Linux  & \texttt{elf64}   & implemented; targeted tests  \\
i386 Linux    & \texttt{elf32}   & implemented; kernel examples \\
\bottomrule
\end{tabular}
\end{table}

\subsection{i386 ABI Notes}
\label{sec:i386-abi}

\par The i386 backend follows a simple C-style stack ABI for exported functions and extern declarations. Primitive arguments are passed through the caller's stack frame, return values use the target return-register convention, and \texttt{extern} declarations describe symbols implemented outside the CPL module. Global CPL functions are emitted as externally visible symbols and can be called from C support code when matching prototypes are used.
\par
\begin{figure*}[t]
\centering
\begin{tikzpicture}[
    >=Latex,
    cell/.style={draw, text width=3.55cm, minimum height=0.56cm, inner xsep=0pt, align=center, font=\footnotesize},
    note/.style={font=\footnotesize, align=center},
    marker/.style={font=\footnotesize, anchor=east},
    arrow/.style={-{Latex[length=1.6mm,width=1.2mm]}, semithick}
]

\node[note] at (0, 0.85) {Normal i386 function after prologue};
\node[cell, fill=black!4] (n0) at (0, 0) {\texttt{[ebp + 12]} second argument};
\node[cell, fill=black!4, below=0cm of n0] (n1) {\texttt{[ebp + 8]} first argument};
\node[cell, fill=black!8, below=0cm of n1] (n2) {\texttt{[ebp + 4]} return address};
\node[cell, fill=black!12, below=0cm of n2] (n3) {\texttt{[ebp]} saved \texttt{ebp}};
\node[cell, fill=black!2, below=0cm of n3] (n4) {locals / spills};
\node[marker] at ([xshift=-0.62cm]n3.west) {\texttt{ebp}};
\draw[arrow] ([xshift=-0.56cm]n3.west) -- ([xshift=-0.08cm]n3.west);
\node[marker] at ([xshift=-0.62cm]n4.west) {\texttt{esp}};
\draw[arrow] ([xshift=-0.56cm]n4.west) -- ([xshift=-0.08cm]n4.west);

\node[note] at (5.3, 0.85) {Naked i386 function at entry};
\node[cell, fill=black!4] (k0) at (5.3, 0) {\texttt{[esp + 8]} second argument};
\node[cell, fill=black!4, below=0cm of k0] (k1) {\texttt{[esp + 4]} first argument};
\node[cell, fill=black!8, below=0cm of k1] (k2) {\texttt{[esp]} return address};
\node[cell, fill=black!2, below=0cm of k2] (k3) {caller stack};
\node[marker] at ([xshift=-0.62cm]k2.west) {\texttt{esp}};
\draw[arrow] ([xshift=-0.56cm]k2.west) -- ([xshift=-0.08cm]k2.west);
\end{tikzpicture}%
\caption{i386 stack view for ordinary and \texttt{naked} CPL functions. In this target, ordinary functions use 32-bit registers such as \texttt{eax}, \texttt{ebx}, \texttt{ecx}, \texttt{edx}, \texttt{esp}, and \texttt{ebp}; without \texttt{push ebp; mov ebp, esp}, the backend addresses arguments from \texttt{esp} rather than from \texttt{ebp}.}
\label{fig:i386-naked-stack}
\end{figure*}
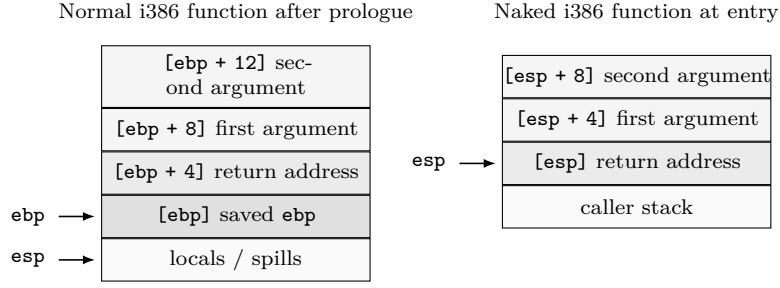

 The \texttt{naked} annotation is the main exception to the ordinary function-frame shape. In a normal i386 function, the compiler may establish an \texttt{ebp}-based frame: \texttt{[ebp + 4]} contains the return address, \texttt{[ebp + 8]} the first argument, and \texttt{[ebp + 12]} the second. A naked function suppresses the generated prologue and epilogue, so the corresponding locations are \texttt{[esp]}, \texttt{[esp + 4]}, and \texttt{[esp + 8]}. Figure~\ref{fig:i386-naked-stack} presents this difference.

\begin{lstlisting}[language=CPL,caption={A naked i386 helper whose arguments are addressed from \texttt{esp}.},label={lst:i386-naked-abi}]
@[nosection] @[naked]
glob function i386_switch2user(u32 eip, u32 esp) -> i0 {
    asm(eip, esp) {
        "cli",
        "push 0x23",
        "push %1",
        "push 0x202",
        "push 0x1b",
        "push %0",
        "iretd"
    }
}
\end{lstlisting}

\par For Listing~\ref{lst:i386-naked-abi}, the backend must materialize the CPL parameters from \texttt{[esp + 4]} and \texttt{[esp + 8]} before expanding inline-assembly placeholders. Once the assembly body pushes interrupt-frame words, \texttt{esp} changes normally; subsequent hand-written instructions must account for that movement.

\section{Static Diagnostics}
\par CPL contains a two-level diagnostic subsystem. The first level operates on the AST and targets source-structure issues such as read-only variable updates, declaration problems, return mismatches, wrong argument counts, incompatible argument types, unused return values, illegal array access, duplicated branches, invalid function names, dead code, lossy implicit conversions, inefficient infinite loops, invalid \texttt{exit} usage, break statements without legal targets, invalid use of \texttt{i0}, unused expressions, invalid references, and suspicious alignment requests.
\par The second level is intentionally placed after SSA construction~\cite{cytron1991ssa}. At this point, the compiler has explicit use-definition chains and phi nodes, so the checker can reason about values and path conditions in a more structured form than in the original HIR. Z3-backed symbolic execution~\cite{deMoura2008z3} is applied to SSA-form HIR to detect definite null dereferences, null values passed to dereferencing functions, possible null dereferences under path-dependent conditions, and constant branches. This ordering is important: the solver-facing analysis works on SSA-form HIR rather than on pre-SSA HIR, because SSA simplifies symbolic value tracking and makes the generated constraints easier to relate to compiler variables.
\par The analysis is diagnostic rather than protective. It does not implement memory safety, ownership, or aliasing restrictions, and it does not prevent undefined behavior caused by dangling pointers, unsafe inline assembly, invalid external interfaces, or target-specific misuse.

\begin{lstlisting}[language=CPL,caption={Example that triggers SSA-level null-dereference diagnostics.}]
start() {
    function foo(ptr i32 p) -> i32 {
        return dref p; :/ Dereference of NULL /:
    }
    ptr i32 a = 0;
    foo(a);
}
\end{lstlisting}

\par A representative diagnostic sequence for this example reports both that the return value of \texttt{foo(a)} is ignored and that \texttt{p} may be dereferenced while equal to null. More complex inputs additionally trigger path-sensitive reports, for example, when one branch makes a pointer definitely null while another makes it only conditionally null.

\subsection{Z3-Backed Symbolic Layer}
\par The SSA-level checker is implemented as a symbolic layer over the HIR control-flow graph. The analyzer prepares HIR expressions as Z3 formulas, preserves symbolic names for compiler variables, and augments path conditions with phi-node constraints when control-flow edges enter SSA merge points. This design makes it possible to ask local reachability and value questions over the same representation that the optimizer sees.
\par The wrapper accepts either parsed JSON or textual HIR dumps, can select a particular function, builds a CFG, and then dispatches solver-backed queries. Two query modes are especially useful for development: \texttt{label}, which checks whether a label can be reached under satisfiable path conditions, and \texttt{var-eq}, which checks whether a selected variable can, must, or cannot equal a given value. The wrapper also supports initial assignments, pointer-width configuration, strict parsing modes, and cached parsing/analysis artifacts. These features keep the solver path useful for compiler diagnostics, as well as for investigating IR examples during pass development.

\begin{lstlisting}[caption={Representative interface to the symbolic query layer.}]
analyze-hir --function entry --pointer-width 64 variable-equals temporary_7 0
\end{lstlisting}

\par This subsystem should be understood as an experimental symbolic analysis component, not as a formal verification framework. Z3 is used to answer bounded path and value questions that are helpful for diagnostics such as null-dereference detection and unreachable-code reasoning. It does not by itself provide a complete semantic proof of CPL programs or of compiler transformations~\cite{deMoura2008z3}.

\subsection{Scalability}

\par The present evaluation does not isolate solver time or characterize its growth with function size and path count. It also does not use a labeled defect corpus from which false-positive and false-negative rates could be computed. Consequently, the evaluation of the symbolic diagnostic layer remains architectural: SSA-form HIR can drive path-conditioned Z3 queries, but the speed and diagnostic accuracy of this design remain unestablished. A quantitative answer requires functions stratified by HIR size and path count, seeded null defects with known ground truth, and separate measurements of solver and total compilation time.

\section{Optimization Passes}

\par Table~\ref{tab:optimizations} summarizes the passes included in the evaluated optimization profiles. The driver uses \texttt{-O0} as the default, enables LICM, constant optimization, and peephole cleanup at \texttt{-O2}, and additionally enables LIR copy propagation and tail-recursion elimination at \texttt{-O3}. Experimental function inlining is excluded from the evaluated pass set because known correctness defects prevent an interpretable performance claim.

\begin{table*}[t]
\centering
\caption{Optimization passes implemented in the CPL compiler prototype.}
\label{tab:optimizations}
\begin{tabular}{p{0.24\linewidth}p{0.50\linewidth}p{0.16\linewidth}}
\toprule
Pass & IR level and role & Level \\
\midrule
LICM & Works on SSA HIR; hoists loop-invariant expressions after loop canonicalization. & \texttt{-O2} and above \\
Peephole & Works on selected LIR after instruction selection, register allocation, memory selection, and caller-saving insertion. & \texttt{-O2} and above \\
DAG rebuild and sparse constant propagation & Works on HIR via DAG generation and CFG rebuild before lowering to LIR. & \texttt{-O2} and above \\
Dead-function elimination & Builds and applies a call graph before later HIR transformations. & default pipeline \\
LIR copy propagation and unused-variable dropping & Rewrites propagated LIR operands and removes unused variable writes. & \texttt{-O3} \\
Tail-recursion elimination & Works on HIR and rewrites tail self-calls to loops before SSA construction. & \texttt{-O3} \\
Hot/cold placement & Implemented during AST-to-HIR generation as annotation-driven branch layout. & default pipeline \\
\bottomrule
\end{tabular}
\end{table*}
\subsection{Loop-Invariant Code Motion}

\par LICM operates on SSA High IR. A loop body that repeatedly computes a constant expression such as \texttt{10 + 10} can have this computation moved before the loop, with phi nodes preserving loop-carried values.

\Needspace{14\baselineskip}
\begin{lstlisting}[language=CPL,caption={LICM source pattern.}]
start() {
    i32 d = 0;
    loop {
        i32 c = 10 + 10;
        d += c;
    }
}
\end{lstlisting}

\subsection{Peephole Optimization}

\par Peephole optimization runs late, after instruction selection and register allocation. It removes redundant register-to-itself moves, simplifies decrement-and-branch sequences, and can replace zero materialization with xor-style idioms where applicable.

\begin{lstlisting}[caption={Peephole effect on a counted empty loop.}]
Before:
    rax = rcx;
    rax = rax - 1;
    rcx = rax;
    rdx = rcx;
    cmp rdx, 0;
    je lb11;
    jne lb9;

After:
    rcx = rcx - 1;
    jne lb10;
\end{lstlisting}

\subsection{PTRN: A Peephole Pattern DSL}

In addition to handwritten rewrites, the late peephole pass also uses \texttt{PTRN}, a small domain-specific language for generating peephole optimization patterns. The motivation is practical: low-level instruction cleanup often consists of many small local rewrites, and encoding them in C by hand makes the pass difficult to extend. PTRN provides a declarative syntax for matching short Low LIR instruction sequences and replacing them with simpler or more efficient sequences.
\par A PTRN file consists of rewrite rules separated by \texttt{/}. The left-hand side describes a sequence to match; the right-hand side after \texttt{->} describes the replacement. Pattern objects abstract over concrete Low LIR operands: \texttt{areg\_N} tracks a repeated abstract register, \texttt{aconst\_N} tracks a repeated abstract constant, \texttt{mem\_N} matches memory locations, and \texttt{obj} can match an arbitrary object. Conditions such as \texttt{[if:equals]}, \texttt{[if:arg2:zero]}, or \texttt{[if:arg2:mod2]} restrict a match, while actions can transform matched constants in the replacement.

\begin{lstlisting}[language=PTRN,caption={Examples of PTRN peephole rules.}]
; Remove a jump that immediately targets the next label.
jmp label_1
mklb label_1
->
mklb label_1
/

; Remove redundant self-copy.
mov obj_1, obj_2 [if:equals]
->
delete
/

; Prefer xor-zeroing over moving an immediate zero.
mov areg_1, const 0
->
xor areg_1, areg_1
/
\end{lstlisting}

\par The generated C code is then used by the low-level peephole pass. This keeps the compiler backend extensible: adding a new local simplification can be implemented by adding a pattern rule rather than modifying the peephole engine directly. The approach is intentionally modest; PTRN does not replace global optimization or data-flow analysis. Its role is to make late instruction cleanup explicit, testable, and easier to maintain.
\par This article does not attribute an independent performance improvement to PTRN because the available benchmark data measures the complete optimization pipeline. PTRN is therefore evaluated here as an implementation mechanism for declarative and generated local rewrites, rather than as an isolated source of speedup.

\subsection{Excluded Experimental Passes}

\par The implementation contains heuristic and model-guided function-inlining experiments. They are not part of the evaluated \texttt{-O0}/\texttt{-O2}/\texttt{-O3} evidence in this article. Known call-site rewriting defects and the absence of model, dataset, and accuracy documentation make performance results involving these modes not interpretable. They are therefore treated as implementation prototypes rather than research results.

\section{Examples and Use Cases}

\par The evaluated examples exercise core low-level features rather than large application workloads. Listings are shortened to isolate the language mechanisms relevant to each case.

\begin{table}[htbp]
\centering
\caption{Representative CPL programs.}
\begin{tabular}{p{0.25\linewidth}p{0.62\linewidth}}
\toprule
Program & What it demonstrates \\
\midrule
Hello World & Strings, pointers to string literals, \texttt{strlen}, inline assembly, direct syscall use, and entry-point termination. \\
CRC8 & Global arrays, indexed memory access, loops, pointer arguments, integer operations, and exit-code calculation. \\
Brainfuck interpreter & Arrays, argument access, switch dispatch, \texttt{no\_fall}, \texttt{straight}, loops, function calls, and byte-level tape manipulation. \\
Memory and file helpers & Header-style declarations, syscall wrappers, pointer arithmetic, and low-level I/O abstractions. \\
OS kernel helpers & i386 extern boundaries, interrupt-related routines, keyboard-driver logic, and C ABI integration. \\
Multiboot kernel entry & Packed header layout, aligned sections, linker-visible entry points, register binding, stack construction, and transfer to a higher-level kernel routine. \\
\bottomrule
\end{tabular}
\end{table}

\par Listing~\ref{lst:hello} presents a compact Hello World excerpt using a direct syscall wrapper. The complete version includes the full \texttt{strlen} implementation and register preservation sequence.

\begin{lstlisting}[language=CPL,caption={Shortened Hello World with explicit syscall wrapper.},label={lst:hello}]
function puts(ptr i8 s) -> i0 {
    asm (s, strlen(s)) {
        "mov rax, 33554436",
        "mov rdi, 1",
        "mov rsi, %0",
        "mov rdx, %1",
        "syscall"
    }
}
\end{lstlisting}

\subsection{Operating-System Code on i386}

\par A practical use case for CPL is small freestanding routines used by an operating-system kernel. The i386 backend was added to support this style of code: globally visible routines can be emitted as NASM-compatible 32-bit assembly, while \texttt{extern} declarations make it possible to integrate CPL modules with C kernel code.

\par Listing~\ref{lst:i386-keyboard} presents a shortened PS/2 keyboard driver excerpt. The complete example additionally contains scancode tables, polling helpers, initialization code, and an exported C boundary.

\begin{lstlisting}[language=CPL,caption={Shortened i386 PS/2 keyboard driver excerpt using C externs.},label={lst:i386-keyboard}]
extern function i386_inb(u16 port) -> i8;
extern function i386_outb(u16 port, u8 data) -> i0;
extern function i386_irq_registerHandler(i32 irq, ptr i0 handler) -> i0;

glob arr _key_pressed[128, u8] = { 0 };

function i386_keyboard_handler(ptr i0 _) -> i0 {
    i8 character = i386_inb(0x60);
    if character < 0 || character >= 128; return;
    _key_pressed[character as i32] = 1;
}
\end{lstlisting}

\par Imported and exported symbols form ordinary low-level ABI boundaries. In this organization, CPL expresses driver control flow and data structures, while C or platform support code supplies target-specific primitives.

\subsection{Multiboot Kernel Bootstrap}
\label{sec:multiboot}

\par A second i386 use case is replacement of a conventional assembly bootstrap with a mostly typed CPL translation unit. Multiboot version 0.6.96 requires a 32-bit-aligned header within the first 8192 bytes of the operating-system image. The header begins with the magic value \texttt{0x1BADB002}; its magic, flags, and checksum fields must sum to zero modulo \(2^{32}\). At transfer of control, \texttt{EAX} contains the Multiboot boot-loader magic, \texttt{EBX} points to the boot-information structure, and the operating system must establish its own stack~\cite{multiboot096}.
\par Listing~\ref{lst:multiboot-cpl} expresses these requirements through packed containers, explicit section placement and alignment, a linker-visible entry symbol, and register-bound source variables. Inline assembly remains necessary for instructions whose effects are not represented by ordinary CPL expressions, including loading \texttt{esp}, disabling interrupts, and halting the processor. The header layout, static storage, and call into the kernel routine remain visible to the type system and normal compiler pipeline.

\begin{lstlisting}[language=CPL,caption={Abridged Multiboot-compatible i386 entry unit.},label={lst:multiboot-cpl},basicstyle=\ttfamily\tiny]
@[align(1)]
container multiboot_header {
    u32 magic;
    u32 flags;
    u32 checksum;
    arr reserved[5, u32];
    u32 mode;
    u32 width;
    u32 height;
    u32 depth;
}

@[section(".multiboot", 4)]
glob multiboot_header header = { 0x1BADB002, 7, 3830599671, 0, 0, 0, 0, 0, 0, 640, 480, 32 };

container kernel_stack {
    arr storage[16384, u8];
}

@[section(".bss", 16)]
glob kernel_stack stack;

@[section(".text")]
function kmain(u32 info, u32 magic, u32 stack_top) -> i0;

@[section(".text")]
@[entry("_start")]
@[naked]
function main() -> i0 {
    @[register(4)] u32 magic;
    @[register(5)] u32 info;

    asm(magic, info) {
        "mov %0, eax",
        "mov %1, ebx"
    }
    asm(ref stack + sizeof(kernel_stack)) {
        "mov esp, %0",
        "cli",
        "xor ebp, ebp"
    }

    u32 stack_top = 0;
    asm(stack_top) { "mov %0, esp" }
    kmain(info, magic, stack_top);
    asm() { ".hang: hlt", "jmp .hang" }
}
\end{lstlisting}

\par The case study does not eliminate assembly semantically; privileged and register-specific operations remain explicit. Instead, it narrows handwritten assembly to the target operations that require it, while representing binary layout, symbol placement, initialization, and control transfer in a typed source language. This division is useful for boot code because layout constraints remain auditable without forcing the complete entry path into an untyped assembly file.

\section{Testing Infrastructure}

\par The compiler is tested with a phase-oriented framework rather than only through end-to-end examples. The central principle is phase observability: tests can inspect preprocessing, tokenization, AST construction, semantic analysis, HIR generation, SSA construction, DAG construction, constant-folding analysis, LIR construction, instruction selection, register allocation, peephole optimization, assembly generation, or assembly-level constant folding. This makes regressions easier to localize than in a purely black-box compiler test.
\par CPL tests use an \texttt{OUTPUT} oracle embedded in source files. The oracle can match runtime output, exit codes, and selected compiler dumps. It also supports tolerant matching for unstable compiler-generated identifiers, so tests can focus on semantically important output instead of incidental temporary names. Runtime tests can run generated assembly and can use multiple argument cases inside one file.
\par Two flags are especially important for this workflow. \texttt{BUG} marks an expected failing test and keeps known defects visible without making the entire test run unusable. \texttt{LEAK\_TRACE} enables memory-operation logging for leak localization in compiler-internal tests. Together with phase observability, these mechanisms support regression tracking across frontend, middle-end, backend, and runtime behavior.

\begin{table*}[t]
\centering
\footnotesize
\caption{Compiler phases exposed by the integration testing framework.}
\label{tab:testing-modules}
\begin{tabular}{llll}
\toprule
Frontend & HIR / SSA & LIR / backend & Assembly \\
\midrule
\texttt{preproc} & \texttt{hir} & \texttt{lir} & \texttt{asm} \\
\texttt{prep} & \texttt{hir\_ssa} & \texttt{lir\_constfold} & \texttt{asm\_constfold} \\
\texttt{ast} & \texttt{hir\_dag} & \texttt{lir\_selector} &  \\
\texttt{sem} & \texttt{hir\_constfold} & \texttt{lir\_instplan} &  \\
 &  & \texttt{lir\_regalloc} &  \\
 &  & \texttt{lir\_peephole} &  \\
\bottomrule
\end{tabular}
\end{table*}

\par The regression corpus covers simple output programs, counted loops, CRC-style table traversal, arithmetic and function-call kernels, Fibonacci, switch dispatch, and a larger Brainfuck interpreter. It also includes OS-oriented i386 examples. This testing setup is not a proof of correctness, but it checks internal forms and executable behavior and provides a practical mechanism for detecting pass-level regressions.

\section{Experimental Evaluation}
\label{sec:evaluation}

\par The evaluation combines a qualitative systems case study with a microbenchmark study. Table~\ref{tab:evaluation-evidence} summarizes the evidence for the evaluated aspects and separates measured behavior from properties that are only supported by the implementation design.

\begin{table*}[t]
\centering
\caption{Evaluation evidence and limitations for the evaluated aspects.}
\label{tab:evaluation-evidence}
\begin{tabular}{p{0.08\linewidth}p{0.30\linewidth}p{0.35\linewidth}p{0.17\linewidth}}
\toprule
Aspect & Evidence & Supported conclusion & Status \\
\midrule
Systems suitability & Multiboot entry unit and i386 kernel helpers & Typed CPL represents layout, sections, symbols, stack storage, and higher-level control transfer; machine-register and privileged operations still require assembly. & partially answered \\
Runtime performance & Seven ten-run microbenchmarks against GCC and Clang on Linux x86\_64 and i386 & CPL is close on the empty counted loop, but mature C optimizers are substantially faster on arithmetic, branch, call, table, and pointer-string kernels. i386 increases pressure on CPL's lowering and register allocation. & preliminary answer \\
Symbolic diagnostics & Implemented SSA-to-Z3 query layer and diagnostic examples & The representation supports path-conditioned null and reachability queries; precision and cost are unknown. & architectural only \\
Optimization effects & \texttt{-O0}/\texttt{-O3} aggregate comparison & Optimization profiles affect runtime, but individual pass contributions and code-size effects cannot be identified. & unanswered \\
\bottomrule
\end{tabular}
\end{table*}

\par The chosen kernels expose loop overhead, arithmetic recurrence, predictable branching, function-call overhead, global table traversal, pointer-heavy string traversal, and a short Fibonacci dependency chain. They are useful for identifying workload-specific behavior but are not a substitute for a standard benchmark suite or kernel-level performance comparison.

\par The microbenchmarks compare CPL with GCC and Clang C baselines on x86\_64 Linux and i386 Linux targets. GCC, Clang, and CPL are tested at \texttt{-O0} and \texttt{-O3}. CPL sources are emitted as NASM assembly, assembled with \texttt{nasm}, and linked with \texttt{ld -e \_main}. The C baselines use freestanding \texttt{\_main} entry points to avoid libc and CRT-startup differences. Each reported runtime is the arithmetic mean of ten executions of one produced binary. The harness and raw repetitions are stored in \texttt{specs/run\_compiler\_microbenches.py} and \texttt{specs/compiler\_microbench\_results.json}. The empty-loop benchmark is a loop-overhead test: the C baseline uses \texttt{asm volatile} to prevent deletion of the loop.

\subsection{Benchmarked Kernels}

\par Table~\ref{tab:bench-kernels} lists the benchmark snippets. The benchmark harness extracts the executable CPL block from each file before the \texttt{OUTPUT} test-runner section.

\begin{table*}[t]
\centering
\caption{Microbenchmark kernels.}
\label{tab:bench-kernels}
\begin{tabular}{lll}
    \toprule
    Kernel & CPL source & Main behavior \\
    \midrule
    Empty counted loop & \texttt{02\_count\_to\_billion.cpl} & preserved one-billion-iteration counter loop \\
    Arithmetic recurrence & \texttt{04\_arith\_mix.cpl} & repeated byte-masked integer recurrence \\
    Hot branch loop & \texttt{05\_branch\_hot.cpl} & predictable branch in a 200M-iteration loop \\
    Hot function call & \texttt{06\_function\_call\_hot.cpl} & small function called in a 100M-iteration loop \\
    Global table traversal & \texttt{07\_table\_sum\_hot.cpl} & repeated scan of a global byte table \\
    Pointer string scan & \texttt{08\_pointer\_string\_scan.cpl} & repeated pointer walk over a string literal \\
    Fibonacci recurrence & \texttt{09\_fibonacci.cpl} & short loop-carried arithmetic dependency \\
    \bottomrule
\end{tabular}
\end{table*}

\subsection{Optimized Runtimes}

\par
\begin{figure*}[t]
\centering
\begin{tikzpicture}
\begin{axis}[
    width=0.92\textwidth,
    height=0.32\textheight,
    ybar,
    bar width=8pt,
    ymin=0,
    ylabel={Runtime (s)},
    symbolic x coords={Empty,Arith,Branch,Call,Table,String,Fib},
    xtick=data,
    x tick label style={rotate=18,anchor=east,font=\scriptsize},
    legend style={at={(0.5,-0.20)},anchor=north,legend columns=3},
    ymajorgrids=true,
    grid style={dashed,gray!30},
    enlarge x limits=0.18,
    nodes near coords,
    nodes near coords align={vertical},
    point meta=rawy,
    every node near coord/.append style={font=\scriptsize}
]
\addplot coordinates {(Empty,0.261887) (Arith,0.319541) (Branch,0.263617) (Call,0.151619) (Table,0.091685) (String,0.084198) (Fib,0.000938)};
\addplot coordinates {(Empty,0.260706) (Arith,0.264630) (Branch,0.080978) (Call,0.105209) (Table,0.002133) (String,0.004427) (Fib,0.000467)};
\addplot coordinates {(Empty,0.261217) (Arith,0.021762) (Branch,0.056354) (Call,0.026664) (Table,0.002328) (String,0.000176) (Fib,0.000470)};
\legend{CPL \texttt{-O3},GCC \texttt{-O3},Clang \texttt{-O3}}
\end{axis}
\end{tikzpicture}
\caption{x86\_64 optimized runtimes for the benchmark kernels. The full x86\_64/i386 table is reported in Table~\ref{tab:benchmarks}.}
\label{fig:bench-optimized}
\end{figure*}
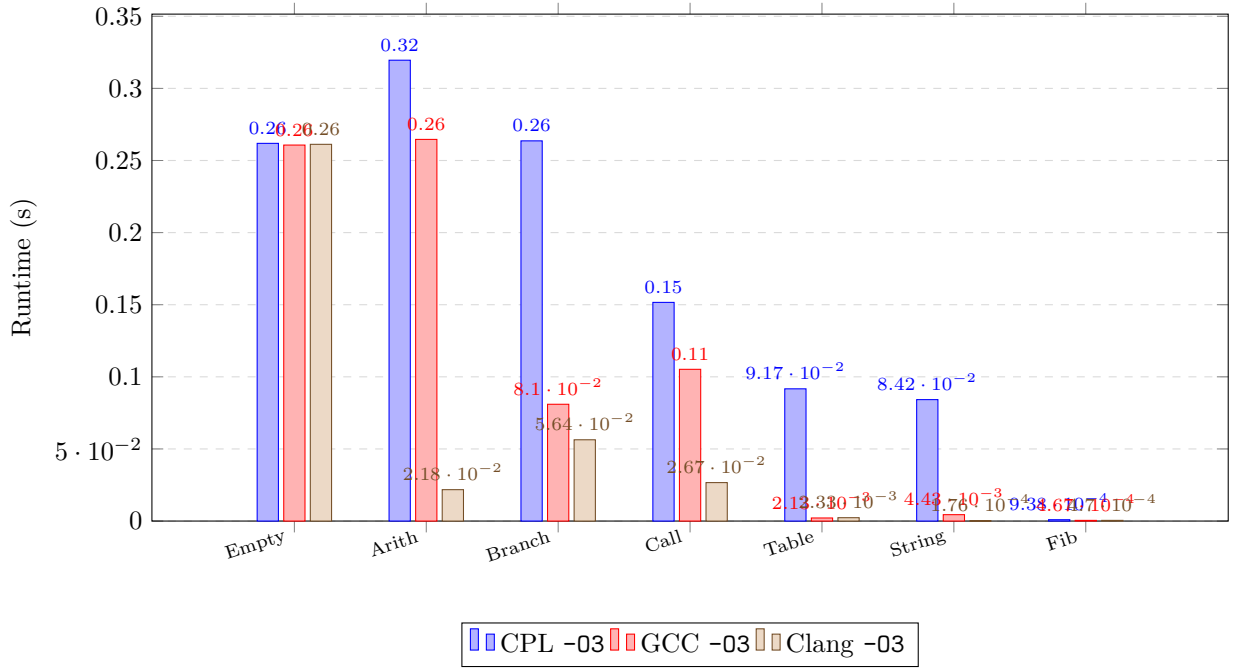

 Table~\ref{tab:benchmarks} reports the optimized and unoptimized runtimes. Figure~\ref{fig:bench-optimized} visualizes the optimized measurements.

\begin{table*}[t]
\centering
\scriptsize
\caption{Mean benchmark runtimes in seconds over ten runs. Lower is better.}
\label{tab:benchmarks}
\begin{tabular}{llrrrrrr}
    \toprule
    Target & Benchmark & CPL \texttt{-O3} & GCC \texttt{-O3} & Clang \texttt{-O3} & CPL \texttt{-O0} & GCC \texttt{-O0} & Clang \texttt{-O0} \\
    \midrule
    x86\_64 & Empty loop       & 0.261887 & 0.260706 & 0.261217 & 0.943493 & 2.732771 & 2.743791 \\
    x86\_64 & Arithmetic       & 0.319541 & 0.264630 & 0.021762 & 0.420121 & 0.472506 & 0.426030 \\
    x86\_64 & Branch           & 0.263617 & 0.080978 & 0.056354 & 0.469712 & 0.522766 & 0.520616 \\
    x86\_64 & Function call    & 0.151619 & 0.105209 & 0.026664 & 0.286096 & 0.286031 & 0.315833 \\
    x86\_64 & Table traversal  & 0.091685 & 0.002133 & 0.002328 & 0.143158 & 0.119041 & 0.140017 \\
    x86\_64 & String scan      & 0.084198 & 0.004427 & 0.000176 & 0.135586 & 0.095225 & 0.106931 \\
    x86\_64 & Fibonacci        & 0.000938 & 0.000467 & 0.000470 & 0.001268 & 0.002113 & 0.002122 \\
    i386 & Empty loop       & 0.261450 & 0.261913 & 0.261629 & 0.951507 & 2.744684 & 2.751128 \\
    i386 & Arithmetic       & 0.474456 & 0.327300 & 0.021911 & 0.540225 & 0.535768 & 0.433954 \\
    i386 & Branch           & 0.381038 & 0.150469 & 0.072008 & 0.464537 & 0.470560 & 0.415575 \\
    i386 & Function call    & 0.330364 & 0.163568 & 0.026601 & 0.399027 & 0.396996 & 0.374836 \\
    i386 & Table traversal  & 0.112236 & 0.009064 & 0.002387 & 0.194468 & 0.121796 & 0.132445 \\
    i386 & String scan      & 0.142074 & 0.030128 & 0.000196 & 0.191890 & 0.094678 & 0.106399 \\
    i386 & Fibonacci        & 0.002809 & 0.000456 & 0.000479 & 0.002066 & 0.002117 & 0.002140 \\
    \bottomrule
\end{tabular}
\end{table*}

\par In the runtime comparison, CPL is effectively tied with GCC and Clang on the preserved empty counted loop on both x86\_64 and i386. On x86\_64, CPL is also within the same millisecond-scale range on Fibonacci, but GCC and Clang are faster on the other optimized kernels. On i386, CPL slows down more strongly on 64-bit arithmetic, function calls, pointer-string traversal, and Fibonacci, which is consistent with higher pressure from 32-bit register allocation and 64-bit operation lowering. The largest gaps remain table traversal and pointer string scanning, where the C compilers appear to perform stronger loop and memory simplification. These explanations remain hypotheses until validated with decoded instruction counts and hardware counters.

\subsection{Effect of Optimization within CPL}

\par
\begin{figure}[t]
\centering
\begin{tikzpicture}
\begin{axis}[
    width=0.96\columnwidth,
    height=0.22\textheight,
    ybar,
    bar width=8pt,
    ymin=0,
    ylabel={Runtime (s)},
    symbolic x coords={Empty,Arith,Branch,Call,Table,String,Fib},
    xtick=data,
    x tick label style={rotate=18,anchor=east,font=\scriptsize},
    legend style={at={(0.5,-0.28)},anchor=north,legend columns=2,font=\scriptsize},
    ymajorgrids=true,
    grid style={dashed,gray!30},
    enlarge x limits=0.25,
    nodes near coords,
    nodes near coords align={vertical},
    point meta=rawy,
    every node near coord/.append style={font=\scriptsize}
]
\addplot coordinates {(Empty,0.943493) (Arith,0.420121) (Branch,0.469712) (Call,0.286096) (Table,0.143158) (String,0.135586) (Fib,0.001268)};
\addplot coordinates {(Empty,0.261887) (Arith,0.319541) (Branch,0.263617) (Call,0.151619) (Table,0.091685) (String,0.084198) (Fib,0.000938)};
\addplot coordinates {(Empty,0.951507) (Arith,0.540225) (Branch,0.464537) (Call,0.399027) (Table,0.194468) (String,0.191890) (Fib,0.002066)};
\addplot coordinates {(Empty,0.261450) (Arith,0.474456) (Branch,0.381038) (Call,0.330364) (Table,0.112236) (String,0.142074) (Fib,0.002809)};
\legend{x86\_64 CPL \texttt{-O0},x86\_64 CPL \texttt{-O3},i386 CPL \texttt{-O0},i386 CPL \texttt{-O3}}
\end{axis}
\end{tikzpicture}
\caption{Optimization effect within CPL.}
\label{fig:bench-cpl-levels}
\end{figure}
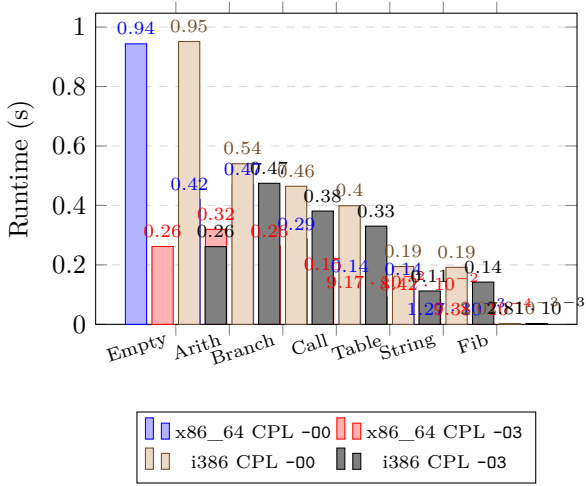

 Figure~\ref{fig:bench-cpl-levels} presents the effect of CPL optimization alone by comparing \texttt{-O0} and \texttt{-O3} on both evaluated architectures. CPL \texttt{-O3} improves most measured kernels, but i386 Fibonacci is a counterexample in these measurements. Within CPL, LICM affects loop-invariant HIR expressions, peephole cleanup removes redundant selected-LIR instructions, and copy propagation reduces low-level temporary traffic after lowering. Because no pass-by-pass ablation was performed, the observed improvement cannot be assigned to an individual optimization.

\par Generated-code size was not measured in object bytes or decoded instruction counts. Source-level or assembly-line counts are omitted because directives, labels, and formatting make them an unreliable proxy for machine-code size.

\section{Limitations and Threats to Validity}

\par The results are subject to the following limitations.

\paragraph{Partial formalization only.}
\par Section~\ref{sec:formal-core} defines typing and reduction rules for a small pointer-and-container core. It does not cover the complete grammar, target annotations, inline assembly, or every control-flow construct, and it is not mechanized. The compiler has no semantic-preservation proof or translation validation.

\paragraph{No memory safety.}
\par The static analyzer can find selected problems, but CPL intentionally does not implement an ownership or borrowing discipline.

\paragraph{Limited target validation.}
\par The most exercised path is x86\_64 Mach-O/NASM. x86\_64 GNU/Linux NASM is implemented but less tested. The i386 GNU/Linux NASM backend has dedicated assembly-generation tests and real operating-system helper examples, but it is newer and has not yet been evaluated with the same breadth as the Mach-O path. Other architecture and system-type options exposed by the driver should be treated as planned or partial support unless backed by dedicated tests.

\paragraph{Excluded function-inlining prototypes.}
\par Heuristic and model-guided inlining modes exist in the implementation but are excluded from the evaluated optimization profiles and contribution claims because their correctness and effectiveness have not been established.

\paragraph{Small benchmark suite.}
\par The benchmark suite consists of microbenchmarks. It does not include large applications, compiler self-hosting, SPEC-style workloads, kernel compilation workloads, broad target comparisons, or enough kernels to characterize general performance.

\paragraph{Single-machine measurements.}
\par Measurements were performed on a single Linux x86\_64 host while targeting both x86\_64 and i386 binaries. Results may vary across processors, operating systems, assemblers, linkers, and timing wrappers.

\paragraph{No cache or memory-hierarchy analysis.}
\par The evaluation does not include cache-miss counters, branch-misprediction counters, memory-bandwidth measurements, or other memory-hierarchy analysis. Pointer-heavy results such as string traversal should therefore be interpreted cautiously.

\paragraph{No differential fuzzing campaign.}
\par The phase-oriented regression suite checks known examples and internal representations, but it does not provide randomized differential validation against a reference interpreter or a semantically equivalent C subset. Compiler reliability beyond the covered tests is therefore not quantified.

\paragraph{Evaluation scope.}
\par The benchmark data should be treated as prototype evidence for the implementation direction rather than as a complete evaluation of every backend and optimization combination.

\paragraph{Inline assembly opacity.}
\par Inline assembly is copied into the output after placeholder substitution and is not optimized by the compiler. This can invalidate assumptions made by surrounding optimization passes if the programmer uses labels, jumps, or unpreserved registers in fragile ways.

\section{Related Work}

\par Table~\ref{tab:related-comparison} compares CPL with projects that represent distinct points in the design space: LLVM~\cite{lattner2004llvm}, QBE~\cite{qbe}, TinyCC~\cite{tinycc}, Zig~\cite{ziglangref}, CompCert~\cite{leroy2009compcert,compcert}, Cogent~\cite{oconnor2016cogent}, and Low*~\cite{protzenko2017lowstar}. The comparison separates language scope, primary objective, low-level access, formal assurance, and backend maturity rather than treating all systems as interchangeable compiler projects.

\begin{table*}[t]
\centering
\footnotesize
\setlength{\tabcolsep}{3pt}
\caption{Positioning of CPL relative to representative systems languages and compiler infrastructures.}
\label{tab:related-comparison}
\begin{tabular}{p{0.10\linewidth}p{0.17\linewidth}p{0.20\linewidth}p{0.17\linewidth}p{0.14\linewidth}p{0.13\linewidth}}
\toprule
Project & Scope & Primary objective & Low-level facilities & Formal assurance & Backend maturity \\
\midrule
LLVM & reusable compiler infrastructure & broad optimization and target support & low-level IR; frontend-dependent source facilities & no general frontend correctness proof & production, many targets \\
QBE & compact SSA backend & simple reusable code generation & low-level IR, not a systems source language & none claimed & small but established backend \\
TinyCC & compact C compiler & fast compilation and C compatibility & C pointers, ABI and system interfaces & none claimed & practical multi-platform C compiler \\
Zig & full systems language & production systems development & explicit memory, ABI, inline assembly, compile-time facilities & language safety checks, no verified compiler claim & production-oriented, broad targets \\
CompCert & C compiler & semantic preservation & C-level systems facilities & machine-checked compiler proof & mature verified target set \\
Cogent/Low* & restricted verified systems languages & proof-oriented low-level software & controlled memory and C interoperability & mechanized semantics and proofs & specialized verified toolchains \\
CPL & compact source language and compiler & inspectable OS/compiler experiments & pointers, layout, sections, entry points, syscalls, inline assembly & partial paper model; no proof & three x86-family formats, uneven validation \\
\bottomrule
\end{tabular}
\end{table*}

\par The closest implementation-scale comparators are QBE and TinyCC, but each leaves a different gap. QBE provides a compact backend rather than a source language with typed OS-facing constructs. TinyCC provides practical C compatibility, but inherits the full complexity and expectations of C and is not organized as a small SSA/SMT experimentation platform. Zig provides substantially stronger language and ecosystem support, while CompCert, Cogent, and Low* provide assurance that CPL does not claim.

\par The resulting niche is narrow: CPL combines an inspectable source language, explicit boot- and kernel-facing controls, a multi-stage optimizing pipeline, and an experimental SSA/SMT diagnostic layer in one compact implementation. This is a research gap only in the sense of an engineering combination, not evidence of superiority. CPL lacks the formal assurance of verified systems, the validation methodology exemplified by Csmith~\cite{yang2011csmith}, and the benchmark breadth and backend maturity of production compilers. The article's contribution is therefore the design and feasibility evidence for that combination.

\par Classes, enums, constructors, destructors, inheritance, and ownership-oriented aggregate models remain outside the core language by design. Containers provide a deliberately small C-like aggregate mechanism, but richer user-defined type systems may be studied separately. This article treats that restriction as part of CPL's deliberate language model rather than as an accidental omission.

\section{Conclusion}

\par The article establishes feasibility of a compact C-like source language that combines OS-facing controls with a conventional optimizing pipeline and SSA/SMT analysis. It does not establish compiler correctness, memory safety, general performance competitiveness, backend maturity, or diagnostic effectiveness. Those claims require the formalization, differential validation, expanded benchmarks, hardware-counter measurements, and labeled diagnostic evaluation specified in the future-work program.

\appendix

\section{Abridged BNF Grammar of CPL}
\label{app:cpl-bnf}

\par This appendix gives an abridged grammar reconstructed from the parser implementation. It documents principal forms rather than a complete formal operational semantics.

\begin{lstlisting}[caption={Abridged CPL grammar.}]
program        ::= top_item*
top_item       ::= annotation* (start_decl | function_decl | container_decl
                 | extern_decl | import_decl | section_decl | align_decl
                 | var_decl | pp_directive | block)

start_decl     ::= "start" "(" param_list? ")" block
function_decl  ::= "function" ident generic_params? "(" param_list? ")"
                   ("->" type)? (";" | block)
extern_decl    ::= "extern" (function_decl | type ident ";")
import_decl    ::= "from" string "import" ident ("," ident)* ";"?

container_decl ::= "container" ident "{" container_item* "}"
container_item ::= annotation* (field_decl | method_decl | pp_directive)
field_decl     ::= storage_mod* annotation* type ident ("=" expr)? stmt_end
method_decl    ::= "function" ident generic_params? "(" param_list? ")"
                   ("->" type)? (";" | block)

param_list     ::= param ("," param)*
param          ::= annotation* ("..." type? ident? | "self" | type annotation* ident ("=" expr)?)
generic_params ::= "<" ident ("," ident)* ">"

storage_mod    ::= "glob" | "ro"
type           ::= prim_type | "ptr" type | "arr" "[" const_len "," type "]"
                 | "(" type_list? ")" "=>" type | ident
prim_type      ::= "f64" | "f32" | "i64" | "i32" | "i16" | "i8"
                 | "u64" | "u32" | "u16" | "u8" | "i0" | "str"
const_len      ::= int | numeric_macro

statement      ::= block | function_decl | align_decl | start_decl
                 | if_stmt | while_stmt | loop_stmt | switch_stmt
                 | return_stmt | exit_stmt | break_stmt | lis_stmt
                 | syscall_stmt | asm_stmt | var_decl | expr ";"
block          ::= "{" statement* "}"
if_stmt        ::= "if" expr ";" statement ("else" statement)?
while_stmt     ::= "while" expr? ";" statement
loop_stmt      ::= "loop" statement
switch_stmt    ::= "switch" expr ";" "{" case_clause* default_clause? "}"
case_clause    ::= "case" literal ";" block
default_clause ::= "default" ";"? block

var_decl       ::= storage_mod* annotation* (array_decl | type ident ("=" expr)? stmt_end)
array_decl     ::= "arr" ident "[" const_len "," type "]" ("=" array_init)? stmt_end
                 | type ident "[" const_len "]" ("=" array_init)? stmt_end

annotation     ::= "@" "[" annotation_text "]"
pp_directive   ::= "#" ("line" | "include" | "define" | "undef"
                 | "ifdef" | "ifndef" | "endif") ... line_end
stmt_end       ::= ";" | end_of_line
\end{lstlisting}

\par The keyword set is: \texttt{start}, \texttt{exit}, \texttt{function}, \texttt{container}, \texttt{return}, \texttt{if}, \texttt{else}, \texttt{while}, \texttt{loop}, \texttt{switch}, \texttt{case}, \texttt{default}, \texttt{glob}, \texttt{ro}, \texttt{dref}, \texttt{ref}, \texttt{ptr}, \texttt{lis}, \texttt{break}, \texttt{extern}, \texttt{from}, \texttt{import}, \texttt{syscall}, \texttt{asm}, \texttt{as}, \texttt{f64}, \texttt{f32}, \texttt{i64}, \texttt{i32}, \texttt{i16}, \texttt{i8}, \texttt{u64}, \texttt{u32}, \texttt{u16}, \texttt{u8}, \texttt{i0}, \texttt{str}, \texttt{arr}, \texttt{not}, \texttt{neg}, \texttt{poparg}, \texttt{sizeof}, \texttt{section}, \texttt{align}, \texttt{line}, \texttt{include}, \texttt{define}, \texttt{undef}, \texttt{ifdef}, \texttt{ifndef}, and \texttt{endif}. Containers may contain scalar fields, pointer fields, and array fields; method-like calls are supported through explicit receiver parameters marked with \texttt{@[self]}. The \texttt{@[like\_c]} annotation is used to document C-ABI-like container padding and layout intent.

\bibliographystyle{plain}
\bibliography{links}

\end{document}